\documentstyle[12pt]{article}

\font\bdi=cmmib10 at 12 pt
\font\twelveof=msym10 at 12pt
\def\bi#1{\hbox{\bdi #1\/}}
\def\xb{\bi{x}}
\def\zb{\bi{z}}
\def\vb{\bi{v}}
\def\wb{\bi{w}}
\def\etab{\bi{\char'21}}
\def\zetab{\bi{\char'20}}
\def\R{\mbox{\twelveof R}}
\def\N{\mbox{\twelveof N}}
\def\case#1#2{{\textstyle{#1\over #2}}}
\newcommand{\sumij}{\sum_{\scriptstyle i,j \atop \scriptstyle i\ne j}}

\begin{document}

\baselineskip=28pt plus 1pt minus 1pt

\begin{center}
{\Large \bf An exactly solvable three-particle problem\\
with three-body interaction}

\bigskip
{\large\bf C.\ Quesne}\footnote{Directeur de recherches FNRS; E-mail address:
cquesne@ulb.ac.be}

{Physique Nucl\'eaire Th\'eorique et Physique Math\'ematique,\\ 
Universit\'e Libre de Bruxelles, Campus de la Plaine CP229,\\ 
Boulevard du Triomphe, B-1050 Brussels, Belgium}

\vspace{3cm}
{\bf Abstract}
\end{center}

The energy spectrum of the three-particle Hamiltonian obtained by replacing the
two-body trigonometric potential of the Sutherland problem by a three-body one of
a similar form is derived. When expressed in appropriate variables, the
corresponding wave functions are shown to be expressible in terms of Jack
polynomials. The exact solvability of the problem with three-body interaction is
explained by a hidden sl(3,\R) symmetry.

\bigskip\noindent
PACS: 03.65.Fd
\newpage
%
%===========================================================================
%
In recent years, the Sutherland one-dimensional $N$-particle
model~\cite{sutherland} and its rational limit, the Calogero
model~\cite{calogero}, have received considerable attention in the literature.
They are indeed relevant to several apparently disparate physical problems, such as
fractional statistics and anyons~\cite{leinaas}, spin chain models~\cite{haldane},
soliton wave propagation~\cite{chen}, two-dimensional non-perturbative quantum
gravity and string theory~\cite{kazakov}, two-dimensional
QCD~\cite{minahan}, quantum chaotic systems and continuous matrix
models~\cite{simons}.\par
%
%-------------------------------------------------------------------------
%
Discovering new exactly solvable problems of a similar kind is of considerable
interest and is therefore a topic of active research (see
e.g.~\cite{yamamoto,khare}).\par
%
%------------------------------------------------------------------------
%
In the present paper, we shall present one such example, corresponding to a
three-particle one-dimensional problem, wherein the particles are assumed
to have equal masses, to move on an interval of length~$\pi/a$, and to
interact via a three-body trigonometric potential. We shall obtain the energy
spectrum and the wave functions of the model when the particles are
distinguishable or when they are indistinguishable and are either bosons or
fermions. In addition, we shall prove that the model exact solvability can be
explained by a hidden sl(3,\R) symmetry.\par
%
%==========================================================================
%
In units wherein $\hbar = 2m = 1$, the Hamiltonian of the problem can be written
as 
\begin{equation}
  H = - \sum_{i=1}^3 \partial_i^2  
       + 3f a^2  \sum_{\scriptstyle i,j,k=1 \atop \scriptstyle i\ne j\ne k\ne i}^3 
       \csc^2  \left(a(x_i+x_j-2x_k)\right), \label{eq:H}
\end{equation}
where $x_i$, $i=1$, 2,~3, $0\le x_i \le \pi/a$, denote the particle coordinates,
$\partial_i \equiv \partial/\partial x_i$, and $f$ is assumed to be such that $-1/4
< f \ne 0$. In the limit where $a\to 0$, the three-body trigonometric potential
in~(\ref{eq:H}) goes over into the three-body inverse square potential for particles
moving on a line that was studied a long time ago by Calogero and
Marchioro~\cite{marchioro}, and, with an additional two-body harmonic potential,
by Wolfes~\cite{wolfes}.\par
%
%-----------------------------------------------------------------------
%
The Hamiltonian is invariant under translations of the centre-of-mass, whose
coordinate will be denoted by $R = (x_1 + x_2 + x_3)/3$. In other words, $H$
commutes with the total momentum $P = -i \sum_{i=1}^3 \partial_i$, which may be
simultaneously diagonalized. It will prove convenient to use two different systems
of relative coordinates, namely $x_{ij} \equiv x_i - x_j$, $i\ne j$, and $y_{ij} \equiv
x_i + x_j - 2x_k$, $i\ne j\ne k\ne i$, where in the latter, we suppressed index~$k$
as it is entirely determined by $i$ and~$j$.\par
%
%-------------------------------------------------------------------------
%
Since for singular potentials crossing is not allowed, in the case of distinguishable
particles the wave functions in different sectors of configuration space are
disconnected, while for indistinguishable particles, they are related by a symmetry
requirement. In the present case, the sector boundaries are determined by
the vanishing of one of the variables~$y_{ij}$. Since $y_{12} + y_{23} + y_{31} =
0$, in a given sector one of the variables~$y_{ij}$ must be of opposite sign to that
of the remaining two. So there are altogether six sectors~\cite{marchioro,
wolfes}, which may be labelled by an index $q=0$, 1, $\ldots,~5$, as follows:
\begin{eqnarray*}
  q=0: \left(y_{12}>0, y_{23}<0, y_{31}<0\right) &\equiv & \left(x_{23}>0, x_{31}<0,
          |x_{12}|<\min(x_{23},-x_{31})\right), \nonumber \\
  q=1: \left(y_{12}>0, y_{23}<0, y_{31}>0\right) &\equiv & \left(x_{12}>0, x_{31}<0,
          |x_{23}|<\min(x_{12},-x_{31})\right),\nonumber \\ 
  q=2: \left(y_{12}<0, y_{23}<0, y_{31}>0\right) &\equiv & \left(x_{12}>0, x_{23}<0,
          |x_{31}|<\min(x_{12},-x_{23})\right), \nonumber \\
  q=3: \left(y_{12}<0, y_{23}>0, y_{31}>0\right) &\equiv & \left(x_{31}>0, x_{23}<0,
          |x_{12}|<\min(x_{31},-x_{23})\right), \nonumber \\
  q=4: \left(y_{12}<0, y_{23}>0, y_{31}<0\right) &\equiv & \left(x_{31}>0, x_{12}<0,
          |x_{23}|<\min(x_{31},-x_{12})\right), \nonumber \\
  q=5: \left(y_{12}>0, y_{23}>0, y_{31}<0\right) &\equiv & \left(x_{23}>0, x_{12}<0,
          |x_{31}|<\min(x_{23},-x_{12})\right). 
\end{eqnarray*}
\par
%
%=========================================================================
%
Let us first assume that the particles are distinguishable and let us restrict the
particle coordinates to a given sector of configuration space. By using the
trigonometric identity
\begin{equation}
  \sum_{\scriptstyle i,j,k=1 \atop \scriptstyle i\ne j\ne k\ne 
       i}^3 \cot(a y_{ij}) \cot(a y_{jk}) = 2,      \label{eq:trigono}
\end{equation}
it is easy to show that the unnormalized ground-state wave function of
Hamiltonian~(\ref{eq:H}) is then given by
\begin{equation}
  \psi_0(\xb) = \prod_{\scriptstyle i,j=1 \atop \scriptstyle i\ne j}^3
  \left|\sin(a y_{ij})\right|^{\lambda},    \label{eq:gswf}
\end{equation}
and corresponds to a vanishing total momentum and to an energy eigenvalue 
$E_0 = 24 a^2 \lambda^2$, where $\lambda \equiv (1 + \sqrt{1 + 4f})/2$ (implying
that $f = \lambda (\lambda - 1)$).\par
%
%------------------------------------------------------------------------
%
As usual in such a type of problem~\cite{sutherland}, the remaining solutions of the
eigenvalue equations $H \psi(\xb) = E \psi(\xb)$, and $P \psi(\xb) = p \psi(\xb)$ can
be found by setting $\psi(\xb) = \psi_0(\xb) \varphi(\xb)$. The
function~$\varphi(\xb)$ satisfies the equations $h \varphi(\xb) =
\epsilon \varphi(\xb)$, and $P \varphi(\xb) = p \varphi(\xb)$, where $\epsilon = E -
E_0$, and the gauge-transformed Hamiltonian $h \equiv (\psi_0(\xb))^{-1} (H - E_0)
\psi_0(\xb)$ can be written as
\begin{equation}
  h = - \sum_{i=1}^3 \partial_i^2  
       - \lambda a  \sum_{\scriptstyle i,j,k=1 \atop \scriptstyle i\ne j\ne k\ne i}^3 
       \cot  \left(a y_{ij}\right) \left(\partial_i + \partial_j - 2 \partial_k\right).
       \label{eq:h}
\end{equation}
\par
%
%-------------------------------------------------------------------------
%
In terms of the new variables $z_i \equiv \exp\left(\frac{2}{3} i a (x_i - 2x_j +
4x_k)\right)$, where $(ijk) = (123)$, $h$ and $P$ become
\begin{equation}
  h = 12 a^2 \left( \sum_i \left(z_i \partial_{z_i}\right)^2 + \lambda
  \sumij \frac{z_i + z_j}{z_i - z_j} z_i \partial_{z_i}\right) - \frac{8}{3} a^2
  \left(\sum_i z_i \partial_{z_i}\right)^2,   \label{eq:h(z)}
\end{equation}
and
\begin{equation}
  P = 2 a \sum_i z_i \partial_{z_i},     \label{eq:P}
\end{equation}
respectively. Equations~(\ref{eq:h(z)}) and~(\ref{eq:P}) bear a resemblance to
corresponding results for the Sutherland potential, with $z_i$ defined in such a
case by $z_i \equiv \exp(2i a x_i)$. This hints at a possibility of expressing the
simultaneous eigenfunctions of $h$ and~$P$ in terms of Jack polynomials as in the
case of the Sutherland potential~\cite{lesage}.\par
%
%-------------------------------------------------------------------------
%
By using Eq.~(11) and theorems~3.1 and~5.1 of Ref.~\cite{stanley}, it is indeed
straightforward to prove that such simultaneous (unnormalized) eigenfunctions are
given by
\begin{equation}
  \varphi_{\{k\}}(\xb) = \exp(6i a q R) J_{\{\mu\}}\left(\zb; \lambda^{-1}\right),
  \label{eq:jack}
\end{equation}
and that there are no further eigenfunctions linearly independent
from~(\ref{eq:jack}). Here $q \in \R$, and $J_{\{\mu\}}\left(\zb;
\lambda^{-1}\right)$ denotes the Jack  (symmetric) polynomial in the variables
$z_i$, $i=1$, 2,~3, corresponding to the parameter~$\lambda^{-1}$, and the
partition $\{\mu\} = \{\mu_1 \mu_2\}$ into not more than two
parts. The associated eigenvalues of $h$ and~$P$ are
\begin{equation}
  \epsilon_{\{ k\}} = 4a^2 \left[3 \sum_i k_i^2 - \case{2}{3} \left(\sum_i k_i\right)
  ^2 - 6 \lambda^2\right],    \label{eq:epsilon}
\end{equation}
and
\begin{equation}
  p_{\{k\}} = 2a \sum_i k_i = 2a \left(\sum_i \mu_i + 3q\right),     \label{eq:p}
\end{equation}
respectively. In Eqs.~(\ref{eq:jack}), (\ref{eq:epsilon}), and~(\ref{eq:p}), $\{ k\} =
\{ k_1 k_2 k_3\}$, where $k_1$, $k_2$, and~$k_3$ are defined by
\begin{equation}
  k_1 = q - \lambda, \qquad k_2 = \mu_2 + q, \qquad k_3 = \mu_1 + q + \lambda.
  \label{eq:k}
\end{equation}
\par
%
%-------------------------------------------------------------------------
% 
The gauge-transformed Hamiltonian~$h$ can be separated into two parts,
describing the centre-of-mass and relative motions respectively, $h = h^{cm} +
h^{rel}$, where $h^{cm} = P^2/3$. As in the case of the Sutherland
potential~\cite{turbiner95}, it is advantageous to write the relative Hamiltonian
$h^{rel}$ in terms of new variables. By setting 
\begin{equation}
  v_i \equiv \exp(-2i a x_{jk}) = z_i \exp(-2iaR)
\end{equation}
for $(ijk) = (123)$, and 
\begin{equation}
  \zeta_1\equiv \sum_i v_i, \qquad \zeta_2 \equiv \sum_{i<j} v_i v_j = \sum_i
  v_i^{-1},      \label{eq:zeta}
\end{equation}
one finds
\begin{equation}
  h^{rel} = 8 a^2 \biggl[ \left(\zeta_1^2 - 3\zeta_2\right) \partial^2_{\zeta_1} 
  + (\zeta_1 \zeta_2 - 9) \partial^2_{\zeta_1\zeta_2} + \left(\zeta_2^2 - 3
  \zeta_1\right) \partial^2_{\zeta_2} + (3\lambda + 1) \left(\zeta_1
  \partial_{\zeta_1} + \zeta_2 \partial_{\zeta_2}\right)\biggr].    \label{eq:hrel}
\end{equation}
\par
%
%------------------------------------------------------------------------
%
The eigenfunctions and eigenvalues of~$h$ can be similarly separated into
centre-of-mass and relative contributions,
\begin{eqnarray}
  \varphi_{\{k\}}(\xb) & = & \varphi_{\{k\}}^{cm}(\xb) \varphi_{\{\mu\}}^{rel}(\xb),
           \\
  \varphi_{\{k\}}^{cm}(\xb) & = & \exp\left(i p_{\{k\}} R\right) = \exp\left[2i a
           \left(\sum_i k_i\right) R\right], \\
  \varphi_{\{\mu\}}^{rel}(\xb) & = & J_{\{\mu\}}\left(\vb;\lambda^{-1}\right) = 
           P_{\{\mu\}}\left(\zetab;\lambda^{-1}\right),   
           \label{eq:eigenfunction} 
\end{eqnarray}
and
\begin{eqnarray}
  \epsilon_{\{k\}} & = & \epsilon_{\{k\}}^{cm} + \epsilon_{\{\mu\}}^{rel}, \\
  \epsilon_{\{k\}}^{cm} & = & \frac{1}{3} p_{\{k\}}^2 = \frac{4}{3} a^2 \left(\sum_i
          k_i\right)^2, \\
  \epsilon_{\{\mu\}}^{rel} & = & 4a^2 \left[3 \sum_i k_i^2 
          - \left(\sum_i k_i\right)^2 - 6 \lambda^2\right].        \label{eq:eigenvalue}
\end{eqnarray}
In Eq.~(\ref{eq:eigenfunction}), $P_{\{\mu\}}\left(\zetab;\lambda^{-1}\right)$ is
the polynomial in $\zeta_1$ and~$\zeta_2$, characterized by the
parameter~$\lambda^{-1}$ and the partition $\{\mu\} = \{\mu_1 \mu_2\}$, that is
obtained from the corresponding Jack polynomial in the variables~$v_i$ by making
the change of variables~(\ref{eq:zeta}). In Eq.~(\ref{eq:eigenvalue}), the
relative-motion energy~$\epsilon_{\{\mu\}}^{rel}$ actually depends only upon the
partition~$\{\mu\}$, and not upon~$q$, and may be written as
\begin{equation}
  \epsilon_{\{\mu\}}^{rel} = 8 a^2 \left(\mu_1^2 - \mu_1 \mu_2 + \mu_2^2 
  + 3 \lambda \mu_1\right),     \label{eq:relenergy}  
\end{equation}
so that $P_{\{\mu\}}\left(\zetab;\lambda^{-1}\right)$ satisfies the eigenvalue
equation
\begin{equation}
  h^{rel} P_{\{\mu\}}\left(\zetab;\lambda^{-1}\right) = \epsilon_{\{\mu\}}^{rel}
  P_{\{\mu\}}\left(\zetab;\lambda^{-1}\right).     \label{eq:relequation}
\end{equation}
\par
%
%-----------------------------------------------------------------------
%
For instance, for $\{\mu\} = \{2\}$, one finds
$J_{\{2\}}\left(\zb;\lambda^{-1}\right) = \sum_i z_i^2 + 2\lambda (\lambda+1)^{-1}
\sum_{i<j} z_i z_j$, and $P_{\{2\}}\left(\zetab;\lambda^{-1}\right) = \zeta_1^2 - 2
(\lambda+1)^{-1} \zeta_2$. For general $\{\mu\}$, one can show that
$P_{\{\mu\}}\left(\zetab;\lambda^{-1}\right)$ belongs to the space
$V_{\mu_1}(\zetab)$, where $V_n(\zetab)$, $n \in \N$, is defined as the space of
polynomials in $\zeta_1$ and~$\zeta_2$ that are of degree less than or equal
to~$n$ (hence, $\dim V_n = (n+1)(n+2)/2$).\par
%
%========================================================================
%
Let us now consider the full configuration space for distinguishable particles.
Let $\varphi^{(q) rel}_{\{\mu\}}(\xb)$ denote the function coinciding with
function~(\ref{eq:eigenfunction}) in sector~$q$, and vanishing in the remaining five
sectors. It is obvious that the six wave functions    
$\varphi^{(q) rel}_{\{\mu\}}(\xb)$, $q=0$, 1, $\ldots$,~5, corresponding to a given
partition~$\{\mu\}$, are associated with the same eigenvalue
$\epsilon^{rel}_{\{\mu\}}$ of~$h^{rel}$. In addition, we note
from~(\ref{eq:relenergy}) that $\epsilon^{rel}_{\{\mu_1 \mu_2\}} =
\epsilon^{rel}_{\{\mu_1, \mu_1-\mu_2\}}$. Hence, for generic (i.e. irrational)
$\lambda$~values, the relative energy spectrum levels characterized by any
partition $\{\mu_1 \mu_2\}$ for which $\mu_1 > 2\mu_2$ have a twelvefold
degeneracy, whereas those for which $\mu_1 = 2\mu_2$ have only a sixfold
degeneracy.\par
%
%-----------------------------------------------------------------------
%
Such degeneracies can be explained by considering the symmetry group
of~$h^{rel}$, which is a group of order~12, obtained by combining the six particle
permutations with the identity and the parity transformation $\Pi: x_{ij} \to -
x_{ij}$. These transformations act both on the variables, and on their domains, i.e.,
the configuration space sectors.\par
%
%------------------------------------------------------------------------
%
The variables~$\zeta_1$ and~$\zeta_2$ remain invariant under any even
permutation, but are interchanged under any odd permutation or the parity
transformation. From Eq.~(\ref{eq:eigenfunction}) and the properties of Jack
polynomials, it results that
\begin{equation}
  P_{\{\mu_1\mu_2\}}\left(\zeta_2,\zeta_1;\lambda^{-1}\right) =
  P_{\{\mu_1,\mu_1-\mu_2\}}\left(\zeta_1,\zeta_2;\lambda^{-1}\right),
\end{equation}
showing that the interchange of $\zeta_1$ and $\zeta_2$ is equivalent to the
replacement of $\{\mu_1\mu_2\}$ by $\{\mu_1,\mu_1-\mu_2\}$.\par
%
%------------------------------------------------------------------------
%
Each $q$-sector is invariant under one odd permutation, but under the remaining
permutations, is changed into the sectors for which $q$ has the same parity. For
instance, $q=0 \to q=0$ under (12), $q=0 \to q=2$ under (23) or (132), and $q=0 \to
q=4$ under (31) or (123). The parity transformation, on the other hand, mixes the
even-$q$-sectors with the odd ones, e.g., $q=0 \to q=3$.\par
%
%-----------------------------------------------------------------------
%
By combining these results, it is now clear that the functions
$\varphi^{(q)rel}_{\{\mu_1\mu_2\}}(\xb)$, and
$\varphi^{(q)rel}_{\{\mu_1,\mu_1-\mu_2\}}(\xb)$ for $q=0$, 1, $\ldots$,~5, and
$\mu_1 > 2\mu_2$, or $\varphi^{(q)rel}_{\{\mu_2,2\mu_2\}}(\xb)$ for $q=0$, 1,
$\ldots$,~5, span the representation space of some irreducible representation of
the relative Hamiltonian symmetry group, as it should be.\par
%
%========================================================================
%
Let us next consider the case of indistinguishable particles, either bosons or
fermions. The only allowed wave functions are then completely symmetrical or
antisymmetrical functions, respectively. The previous discussion shows that they
are given by
\begin{eqnarray}
  \varphi^{(\pm)(e)rel}_{\{\mu_1\mu_2\}}(\xb)  & = &
           \varphi^{(0)rel}_{\{\mu_1\mu_2\}}(\xb) 
           + \varphi^{(2)rel}_{\{\mu_1\mu_2\}}(\xb)
           + \varphi^{(4)rel}_{\{\mu_1\mu_2\}}(\xb) \nonumber \\
  & & \mbox{} \pm \left[\varphi^{(0)rel}_{\{\mu_1,\mu_1-\mu_2\}}(\xb) 
           + \varphi^{(2)rel}_{\{\mu_1,\mu_1-\mu_2\}}(\xb)
           + \varphi^{(4)rel}_{\{\mu_1,\mu_1-\mu_2\}}(\xb)\right], 
\end{eqnarray}
and
\begin{eqnarray}
  \varphi^{(\pm)(o)rel}_{\{\mu_1\mu_2\}}(\xb)  & = &
           \varphi^{(1)rel}_{\{\mu_1\mu_2\}}(\xb) 
           + \varphi^{(3)rel}_{\{\mu_1\mu_2\}}(\xb)
           + \varphi^{(5)rel}_{\{\mu_1\mu_2\}}(\xb) \nonumber \\
  & & \mbox{} \pm \left[\varphi^{(1)rel}_{\{\mu_1,\mu_1-\mu_2\}}(\xb) 
           + \varphi^{(3)rel}_{\{\mu_1,\mu_1-\mu_2\}}(\xb)
           + \varphi^{(5)rel}_{\{\mu_1,\mu_1-\mu_2\}}(\xb)\right], 
\end{eqnarray}
where the upper (resp.\ lower) signs correspond to bosons (resp.\ fermions). The
relative energy spectrum levels are characterized by the partitions $\{\mu_1
\mu_2\}$, such that $\mu_1 \ge 2\mu_2$ for bosons, or $\mu_1 > 2\mu_2$ for
fermions, and they have a residual twofold degeneracy coming from the invariance
of~$h^{rel}$ under~$\Pi$. The linear combinations
$\varphi^{(\pm)(e)rel}_{\{\mu_1\mu_2\}} \pm
\varphi^{(\pm)(o)rel}_{\{\mu_1\mu_2\}}$, and
$\varphi^{(\pm)(e)rel}_{\{\mu_1\mu_2\}} \mp
\varphi^{(\pm)(o)rel}_{\{\mu_1\mu_2\}}$ have a given parity, even or odd,
respectively.\par
%
%=========================================================================
% 
The relative energy spectrum (in appropriate units) and its degeneracies are the
same as for three particles on an interval of length $\pi/a$, interacting via the
two-body potential $\kappa (\kappa-1) a^2 \sum_{i\ne j} \csc^2(a x_{ij})$
whenever the particles are distinguishable, but they are distinct for
indistinguishable particles. This is due to the fact that both the configuration space
sectors, and the variables the relative wave functions depend upon have different
transformation properties under permutations for the two potentials.\par
%
%------------------------------------------------------------------------
%
For the two-body potential, the configuration space sectors may be labelled by an
index $p=0$, 1, $\ldots$,~5, and defined as follows: $p=0$: $\left(x_{12}>0,
x_{23}>0, x_{31}<0\right)$, $p=1$: $\left(x_{12}>0, x_{23}<0, x_{31}<0\right)$, 
$p=2$: $\left(x_{12}>0, x_{23}<0, x_{31}>0\right)$, $p=3$:\linebreak
$\left(x_{12}<0, x_{23}<0, x_{31}>0\right)$, $p=4$: $\left(x_{12}<0, x_{23}>0,
x_{31}>0\right)$, $p=5$: $\left(x_{12}<0, x_{23}>0, x_{31}<0\right)$
\cite{sutherland, marchioro, wolfes}. Any one of them is transformed into the five
remaining sectors under permutations. On the other hand, the $p=0$ and $p=3$, $p=1$
and $p=4$, $p=2$ and $p=5$ sectors are interchanged under~$\Pi$.\par
%
%------------------------------------------------------------------------
%
The relative wave functions in a given sector now assume the
form~\cite{turbiner95}
\begin{equation}
  \varphi^{rel}_{\{\mu\}}(\xb) = J_{\{\mu\}}\left(\wb;\kappa^{-1}\right) =
  P_{\{\mu\}}\left(\etab;\kappa^{-1}\right),   \label{eq:eigenfunctionSuth}
\end{equation}
where $w_i \equiv \exp(-2iay_{jk}/3)$ for $(ijk) = (123)$, and $\eta_1 \equiv
\sum_i w_i$, $\eta_2 \equiv \sum_{i<j} w_i w_j = \sum_i w_i^{-1}$. From their
definition, it results that the variables~$\eta_1$ and~$\eta_2$ remain invariant
under permutations, but are interchanged under~$\Pi$.\par
%
%---------------------------------------------------------------------
%
For bosons or fermions, the relative wave functions for the two-body potential
may therefore be written as
\begin{equation}
  \varphi^{(\pm)rel}_{\{\mu\}}(\xb) = \varphi^{(0)rel}_{\{\mu\}}(\xb) +
  \varphi^{(2)rel}_{\{\mu\}}(\xb) + \varphi^{(4)rel}_{\{\mu\}}(\xb) \pm \left[
  \varphi^{(1)rel}_{\{\mu\}}(\xb) + \varphi^{(3)rel}_{\{\mu\}}(\xb) +
  \varphi^{(5)rel}_{\{\mu\}}(\xb)\right],   \label{eq:BFSuth}     
\end{equation}
where $\varphi^{(p)rel}_{\{\mu\}}(\xb)$ denotes as before the function coinciding
with function~(\ref{eq:eigenfunctionSuth}) in sector~$p$, and vanishing in the
remaining five sectors. In Eq.~(\ref{eq:BFSuth}), $\{\mu\}$ runs over all partitions
into not more than two parts.  The relative energy spectrum levels are
characterized by $\{\mu_1 \mu_2\}$, where $\mu_1 \ge 2\mu_2$ for both bosons
and fermions. Those with $\mu_1 > 2\mu_2$ have a twofold degeneracy coming
from the invariance of~$h^{rel}$ under~$\Pi$. The corresponding even and odd wave
functions are given by $\varphi^{(\pm)rel}_{\{\mu_1\mu_2\}}(\xb) \pm 
\varphi^{(\pm)rel}_{\{\mu_1,\mu_1-\mu_2\}}(\xb)$, and 
$\varphi^{(\pm)rel}_{\{\mu_1\mu_2\}}(\xb) \mp 
\varphi^{(\pm)rel}_{\{\mu_1,\mu_1-\mu_2\}}(\xb)$, respectively. In contrast, the
levels with $\mu_1 = 2\mu_2$ are not degenerate, the corresponding wave function
$\varphi^{(+)rel}_{\{\mu_2,2\mu_2\}}(\xb)$ 
(resp.~$\varphi^{(-)rel}_{\{\mu_2,2\mu_2\}}(\xb)$) being even (resp.~odd).\par
%
%========================================================================
%  
As a final point, we shall now proceed to show that the exact solvability of~$H$,
defined in Eq.~(\ref{eq:H}), or equivalently of~$h^{rel}$, defined in
Eq.~(\ref{eq:hrel}), is due to a hidden sl(3,\R) symmetry. For such purpose, let us
consider the operators
$E_{ij}$,
$i$,~$j=1$, 2,~3, defined by
\begin{eqnarray}
  E_{11} & = & \zeta_1 \partial_{\zeta_1}, \qquad E_{22} = \zeta_2
            \partial_{\zeta_2}, \qquad E_{33} = n - \zeta_1 \partial_{\zeta_1}
            - \zeta_2 \partial_{\zeta_2}, \nonumber \\
  E_{31} & = & \partial_{\zeta_1}, \qquad E_{32} = \partial_{\zeta_2}, \qquad
            E_{21} = \zeta_2 \partial_{\zeta_1}, \qquad E_{12} = \zeta_1
            \partial_{\zeta_2}, \nonumber \\
  E_{13} & = & n \zeta_1 - \zeta_1^2 \partial_{\zeta_1} - \zeta_1 \zeta_2
            \partial_{\zeta_2}, \qquad  E_{23} = n \zeta_2  - \zeta_1 \zeta_2
            \partial_{\zeta_1} - \zeta_2^2 \partial_{\zeta_2},   \label{eq:sl-gen} 
\end{eqnarray}
where $n$ may take any real value. It is clear~\cite{turbiner95,turbiner94} that
they fulfil the gl(3,\R) commutation relations
\begin{equation}
  \left[E_{ij}, E_{kl}\right] = \delta_{kj} E_{il} - \delta_{il} E_{kj}.
  \label{eq:sl-com}
\end{equation}
Since the linear combination $\sum_i E_{ii}$ reduces to a constant,
Eq.~(\ref{eq:sl-gen}) actually provides a representation of the traceless part
sl(3,\R) of gl(3,\R), acting on the space of functions in $\zeta_1$ and~$\zeta_2$.
Whenever $n$~is a non-negative integer, such a representation reduces to a
finite-dimensional one on the space~$V_n(\zetab)$ of polynomials in $\zeta_1$
and~$\zeta_2$ that are of degree less than or equal to~$n$.\par
%
%-------------------------------------------------------------------------
%
It is now straightforward to prove that $h^{rel}$ belongs to the enveloping algebra of
sl(3,\R). It can indeed be rewritten as the following quadratic combination of the
$E_{ij}$'s,
\begin{equation}
  h^{rel} = 8 a^2 \left[E_{11}^2 + E_{11} E_{22} + E_{22}^2 - 3 E_{12} E_{32}
  - 3 E_{21} E_{31} - 9 E_{31} E_{32} + 3 \lambda \left(E_{11} + E_{22}\right)
  \right].     \label{eq:envalg}
\end{equation}
Such an expression is valid for any real $n$~value. Hence, the operator~$h^{rel}$
possesses infinitely many finite-dimensional invariant subspaces $V_n(\zetab)$,
$n=0$, 1, 2,~$\ldots$, and, correspondingly, preserves an infinite flag of spaces,
$V_0(\zetab) \subset V_1(\zetab) \subset V_2(\zetab) \subset \ldots$. In the
basis wherein all spaces $V_n(\zetab)$ are naturally defined, the matrix
representing $h^{rel}$ is therefore triangular, so that $h^{rel}$ is exactly
solvable~\cite{turbiner94}.\par
%
%========================================================================
%
In conclusion, we did prove in the present paper that various results valid for the
$N$-particle Sutherland problem can be extended to the three-particle problem,
wherein the Sutherland two-body trigonometric potential is replaced by a
three-body potential of a similar form. In particular, the wave functions can still
be expressed in terms of Jack polynomials~\cite{lesage}, such a property being
related with the existence of a hidden sl(3,\R) symmetry~\cite{turbiner95}.\par
%
%------------------------------------------------------------------------
%
Although the wave functions of the problem with two-body interaction and of the
present one look similar when expressed in appropriate variables, they are actually
rather different when rewritten in terms of the particle coordinates $x_i$, $i=1$,
2,~3. As a consequence, for indistinguishable particles for which permutations of
the $x_i$'s play a crucial role, the energy spectra of the two problems are distinct.
Comparing other properties of the three-particle system for the two problems might
be an interesting question for future study.\par
%
%-----------------------------------------------------------------------
%
It is not clear yet whether the three-particle problem with both two- and
three-body trigonometric potentials is exactly solvable as its rational
limit~\cite{marchioro,wolfes}. Investigating this point would be of interest
too.\par
%
%-------------------------------------------------------------------------
%
One should also note that the $N$-particle problem with three-body trigonometric
interaction is unlikely to be solvable as the three-particle one since the latter is
connected with the exceptional Lie algebra~G$_2$~\cite{perelomov}. This contrasts
with the case of the $N$-particle problem with two-body trigonometric interaction,
which is related to the Lie algebra~A$_{N-1}$ for any $N=2$, 3,~$\ldots$.\par
%
%-----------------------------------------------------------------------
%
As for the Sutherland potential~\cite{lesage}, one might also consider a
generalized spin-dependent Hamiltonian for particles with internal degrees of
freedom. In such a case, use could be made of the three first-order
differential-difference operators that were recently  introduced in connection with
the Weyl group~D$_6$ of G$_2$~\cite{cq1,cq2}.\par
\newpage
%
%========================================================================
%


\begin{thebibliography}{99}

\bibitem{sutherland} B. Sutherland, Phys. Rev. A {\bf 4}, 2019 (1971); {\bf 5}, 1372
(1972); Phys. Rev. Lett. {\bf 34}, 1083 (1975).

\bibitem{calogero} F. Calogero, J. Math. Phys. {\bf10}, 2191 (1969); {\bf 10}, 2197
(1969); {\bf 12}, 419 (1971).

\bibitem{leinaas} J. M. Leinaas and J. Myrheim, Phys. Rev. B {\bf 37}, 9286 (1988);
A. P. Polychronakos, Nucl. Phys. B {\bf 324}, 597 (1989); Phys. Lett. B {\bf 264}, 362
(1991);
F. D. M. Haldane, Phys. Rev. Lett. {\bf 67}, 937 (1991).

\bibitem{haldane} F. D. M. Haldane, Phys. Rev. Lett. {\bf 60}, 635 (1988); {\bf 66},
1529 (1991);
B. S. Shastry, {\em ibid.} {\bf 60}, 639 (1988).

\bibitem{chen} H. H. Chen, Y. C. Lee, and N. R. Pereira, Phys. Fluids {\bf 22}, 187
(1979).

\bibitem{kazakov} V. A. Kazakov, in {\em Random Surfaces and Quantum
Gravity (Cargese Lectures, 1990)}, edited by O. Alvarez {\em et al} (Plenum, New
York, 1991).

\bibitem{minahan} J. A. Minahan and A. P. Polychronakos, Phys. Lett. B {\bf 326},
288 (1994).

\bibitem{simons} B. D. Simons, P. A. Lee, and B. L. Altshuler, Phys. Rev. Lett. {\bf 70},
4122 (1993); {\bf 72}, 64 (1994).

\bibitem{yamamoto} T. Yamamoto, Phys. Lett. A {\bf 208}, 293 (1995).

\bibitem{khare} A. Khare, J. Phys. A {\bf 29}, L45 (1996).

\bibitem{marchioro} F. Calogero and C. Marchioro, J. Math. Phys. {\bf15}, 1425
(1974).  

\bibitem{wolfes} J. Wolfes, J. Math. Phys. {\bf 15}, 1420 (1974).

\bibitem{lesage} F. Lesage, V. Pasquier, and D. Serban, Nucl. Phys. B {\bf 435}, 585
(1995).

\bibitem{stanley} R. P. Stanley, Adv. Math. {\bf 77}, 76 (1989).

\bibitem{turbiner95} W. R\"uhl and A. Turbiner, Mod. Phys. Lett. A {\bf 10}, 2213
(1995).

\bibitem{turbiner94} A. V. Turbiner, in {\em Lie Algebras, Cohomologies and
New Findings in Quantum Mechanics (Contemporary Mathematics 160)}, edited by N.
Kamran and P. Olver (AMS, Providence, RI, 1994) p. 263.

\bibitem{perelomov}M. A. Olshanetsky and A. M. Perelomov, Phys. Rep. {\bf 94}, 313
(1983).

\bibitem{cq1} C. Quesne, Mod. Phys. Lett. A {\bf 10}, 1323 (1995).

\bibitem{cq2} C. Quesne, Europhys. Lett. {\bf 35}, 407 (1996).

\end {thebibliography}
\end{document}